\documentstyle[amssymb,twocolumn,aps,prl,epsfig]{revtex}

\twocolumn

\begin{document}
\draft
\wideabs{
\author{A.M. Akulshin$^{1,2}$, S. Barreiro$^1$, and A. Lezama$^1$}
\address{$^1$ Instituto de F\'{\i }sica, Facultad de Ingenier\'{\i }a, C. Postal.30.\\
11000,\\
Montevideo, Uruguay\\
$^2$ P.N.Lebedev Physics Institute, 117924 Moscow, Russia}
\date{\today }
\title{Steep anomalous dispersion in coherently prepared Rb vapor}
\maketitle

\begin{abstract}
Steep dispersion of opposite signs in driven degenerate two-level atomic
transitions have been predicted and observed on the D$_2$ line of $^{87}$Rb
in an optically thin vapor cell. The intensity dependence of the anomalous
dispersion has been studied. The maximum observed value of anomalous
dispersion ($dn/d\nu \simeq -6\times $ $10^{-11}Hz^{-1}$) corresponds to a
negative group velocity $V_g\simeq -c/23000$.
\end{abstract}

}

Investigations of coherent effects in resonant media, namely coherent
population trapping (CPT) and electromagnetically induced transparency (EIT) 
\cite{HARRIS,ARIMONDO}, which can dramatically modify the absorptive and
dispersive properties of an atomic vapor, have caused a rebirth of interest
in the problem of light propagation through a dispersive medium. In the last
decade, the study of the dispersive properties of coherently prepared media
was always under attention due to fundamental and practical interest.

An ultra-large index of refraction in coherently prepared resonant gas was
predicted\cite{Scully-old} and a refractive index variation as large as $%
\Delta n\approx 1\times 10^{-4}$ was demonstrated in a dense Rb vapor\cite
{Zibrov-96}. It was also shown that a coherently driven medium exhibits
large dispersion\cite{Harris-92}. A high normal dispersion (up to $dn/d\nu
\simeq 1\times 10^{-11}Hz^{-1}$) was measured on the Cs $D_2$ line in a
vapor cell\cite{Mesh} and in an atomic beam\cite{Muller OptCom}. Recently,
extremely slow light group velocity ($17\ m/s$) associated with normal
dispersion was demonstrated in an ultracold atomic sample\cite{L.Hau}.
However, the same order of magnitude of group velocity ($90\ m/s$) was
observed in a hot dense vapor cell\cite{Sau}.

All these investigations were carried on alkaline atoms where the absorption
is strongly suppressed and dispersion is steep and normal ($D\equiv dn/d\nu
>0$) due to CPT between the two ground state hyperfine levels ($\Lambda $
scheme). However, atomic coherence among Zeeman sublevels belonging to the
same ground-state hyperfine level can led not only to usual EIT, but also to
an absorption enhancement named as {\it electromagnetically induced
absorption }(EIA)\cite{LEZAMA,Alexei}. Since EIT/EIA effects in degenerate
two-level systems can produce a significant variation in the absorption with
subnatural width, one can predict a large absolute value of dispersion in
this case. Notice that at resonance dispersion would be normal ($D>0$) for
EIT and anomalous ($D<0$) for EIA. In both cases the absolute value of the
dispersion can be several orders of magnitude greater than for a linear
medium. In this letter we present the first observation of steep anomalous
and normal dispersion in coherently prepared degenerate two-level atomic
system.

Refractive index and dispersion were analyzed with the model recently used
to study subnatural EIA resonances\cite{Lezama-2}. In this model, two
monochromatic optical fields, a drive field and a weak probe field with
amplitudes $E_d$, $E_p$ and frequencies $\omega _d$, $\omega _p$\
respectively are incident on motionless two-level atoms with resonance
frequency $\omega _0$ and electric dipole moment $\mu $. The atomic levels
are degenerate. The configuration is closed. The spontaneous decay rate\ is $%
\Gamma $. Finite interaction time is described by a relaxation rate $\gamma $
($\gamma \ll \Gamma $). The drive wave Rabi frequency is $\Omega =\mu
E_d/\hbar $ and its saturation parameter $S\equiv 2\Omega ^2/\Gamma ^2$.

\begin{figure}[tbp]
\begin{center}
\mbox{\epsfig{file=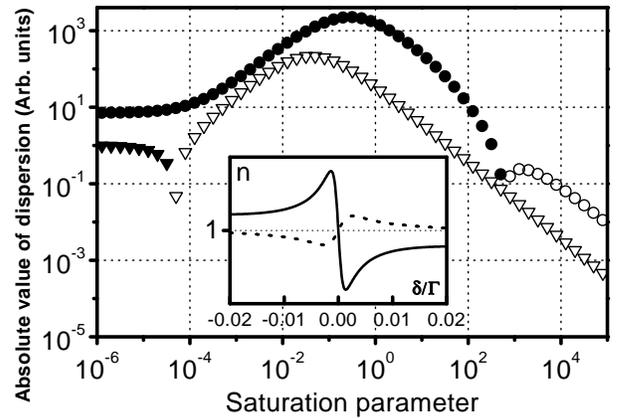,width=3.5in}}
\end{center}
\caption{Calculated dispersion at $\delta=0$\ for the transitions $%
F_g=2\rightarrow F_e=3$ (circles) and $F_g=1\rightarrow F_e=0$ (triangles)\
as a function of the saturation parameter $S$.\ Solid(hollow) points
correspond to negative(positive) dispersion. Inset: Calculated refractive
index as a function of $\delta$\ for linear and orthogonal pump and probe
polarizations for two different transitions and same drive field intensity [$%
\omega _d=\omega _0$, $\Gamma /\gamma $\ $=1000$, $\ $ $S$ $\lesssim 1$].}
\label{calcdisp}
\end{figure}

The calculated refractive index, tested by the probe wave in the presence of
the drive field, as function of the frequency offset $\delta \equiv $ $%
\omega _d-\omega _p$ is presented in the inset of Fig.\ref{calcdisp}. The
shown spectra correspond to the closed transitions in the $D_2$ line of $%
^{87}Rb$. The peak to peak refractive index variation $\Delta n$ is higher
for the anomalous dispersion because the transition $F_g=2\rightarrow F_e=3$%
{\it \ }is stronger than the $F_g=1\rightarrow F_e=0$ transition. The
corresponding absolute value of the anomalous dispersion is also larger.

The values of $D$ and $n$ depend on light intensity, atomic density $N$,
level degeneracy, polarizations, among others parameters. Here we restrict
our attention to the dependence with the drive-field intensity $I_d$ .

Since a general expression for $n$ in driven degenerate two-level systems is
not available at present, we consider as a guide the analytical expression
corresponding to the ideal $\Lambda $ scheme\cite{Muller OptCom}: 
\begin{equation}
n(\delta )=1+\frac 3{8\pi ^2}\lambda ^3N\frac{\Omega ^2\delta }{(\Omega
^2+\Gamma \gamma /2)^2+(\gamma \delta /2)^2}  \label{refract}
\end{equation}
where $\lambda $ is the wavelength of the optical transition. At low
intensity ($\Omega ^2\ll \Gamma \gamma $) $\Delta n$ and $D$ around $\delta
=0$ are growing linearly with $\Omega ^2$. For high intensity ($\Omega ^2\gg
\Gamma \gamma $), $D$ and $\Delta n$ are inversely proportional to $\Omega
^2 $. It can be shown that the maximum for $\Delta n$ and $D$ are reached
for $\Omega ^2=\Gamma \gamma /2$ when the saturation parameter $S=\gamma
/\Gamma $.

Such behavior may be explained in the following way: At low intensity, when
power broadening is not significant (the width of the resonance is
determined by the ground-state relaxation), the amplitude is growing
linearly with intensity, while the width remains almost constant. In this
case, $D$ grows linearly with $I_d$. The dispersion saturates when the
resonance width is determined by power broadening. At high intensity, the
refractive index and the absorption saturate \cite{Lezama-2} while the
resonance width still grows. So, in this region $D$ decreases with intensity.

The calculated intensity dependences of the dispersion at resonance ($\delta
=0$) shown in Fig.\ref{calcdisp} are in qualitative agreement with the
simple analytical expression (Eq.\ref{refract}) for low and high drive
intensity. At very low drive intensity (linear absorption) the dispersion
for the two transitions considered in Fig.\ref{calcdisp} is anomalous. For
higher drive intensity there is absorption enhancement (EIA) for one
transition and absorption reduction (EIT) for the other. The first case
results in anomalous dispersion while the second one corresponds to normal
dispersion. The two curves have a maximum at moderate intensity ($S\lesssim
1 $). At large drive intensities ($S>10^3$) the dispersion on both
transitions is normal and exhibits the same linear asymptotic behavior.

The experiment was realized on the D$_2$ line of $^{87}$Rb in a vapor cell.
We used a phase heterodyne method to measure the refractive index. The idea
of this approach is based on the well-known method of FM spectroscopy and
the two-mode technique\cite{Gubin}. Our method is similar to that used for
dispersion measurements \cite{Muller OptCom} and for Doppler-free
spectroscopy \cite{Dima}. To obtain information about $n$, the phase of a
RF-signal produced by mixing the resonant probe wave with a non-resonant
auxiliary wave (having the same optical path) is compared to an RF reference
produced by the mixing of undisturbed fractions of the probe and the
auxiliary waves. The variation of the phase difference between the two
RF-signals, due to the atomic medium dispersion, is $\delta \Phi
=l(n-1)\omega /c$, where $l$ is the vapor cell length. With these technique,
the influence of acoustic noise is dramatically reduced compared to the
homodyne method based on a Mach-Zehnder interferometer\cite{Zibrov-96,Mesh}.

\begin{figure}[tbp]
\begin{center}
\mbox{\epsfig{file=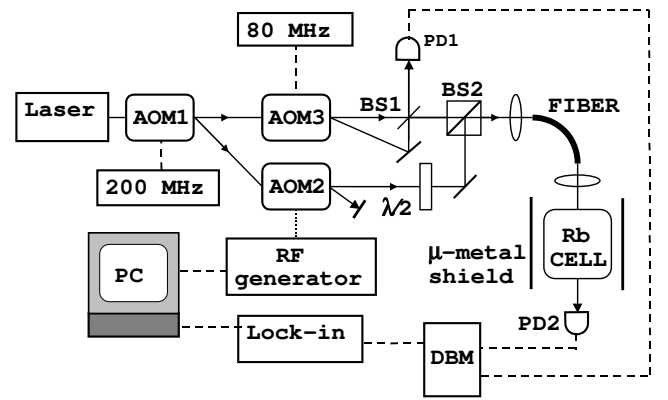,width=3.5in}}
\end{center}
\caption{Experimental setup.}
\label{setup}
\end{figure}

The scheme of the experimental setup is shown in Fig.\ref{setup}. A
single-mode extended cavity diode laser frequency lockable to a Rb saturated
absorption resonance was used. Two mutually coherent waves with tunable
optical frequency offset were obtained by using two acousto optic modulators
(AOM's 1,2) \cite{LEZAMA}. The diffracted output from AOM2, driven by a
tunable RF generator, was used as the drive wave. The two outputs of AOM3
(with fixed frequency offset) were combined on the beamsplitter BS1. One of
this beamsplitter outputs was used as the signal wave. It contains two
frequency components, the resonant probe wave and the non resonant auxiliary
wave. The second output was used to produce the RF-reference on the
photodiode (PD1). Signal and drive waves with orthogonal linear
polarizations were superimposed on the polarization beamsplitter BS2. After
spatial filtering in a 50-cm long single-mode optical fiber, the pump and
signal waves were sent through a 5-cm long Rb cell at room temperature. The
external magnetic field at the cell was reduced to $10\ mG$ level using a $%
\mu $-metal shield. Maximum powers of the drive and the signal waves at the
cell were $0.15\ mW$ and $0.04\ mW$ respectively. A double balanced mixer
(DBM) was used as a phase detector. The voltage-to-phase response of the DBM
was calibrated introducing known delays between the two inputs. A lock-in
amplifier was used for signal processing.

\begin{figure}[tbp]
\begin{center}
\mbox{\epsfig{file=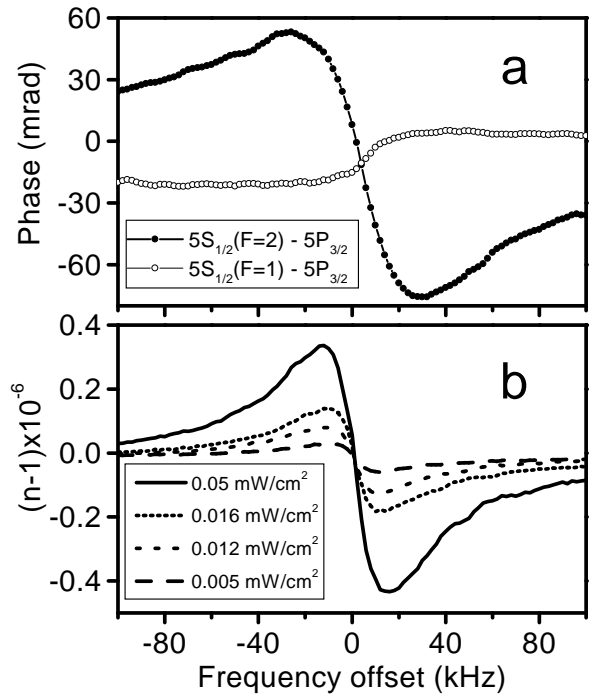,width=3.5in}}
\end{center}
\caption{a) Phase resonances for laser frequencies corresponding to the
transitions $5S_{1/2}(F_g=2)-5P_{3/2}(F_e=3)$ and $%
5S_{1/2}(F_g=1)-5P_{3/2}(F_e=0)$ of $^{87}$Rb (beam diameter: $0.8\ cm$,
drive intensity: $0.3\ mW/cm^2$). b) Refractive index for the transition $%
5S_{1/2}(F_g=2)-5P_{3/2}(F_e=3)$ as a function of $\delta $ for different
values of $I_d$\ and beam diameter $1.6\ cm$.}
\label{expdisp}
\end{figure}

Experimental spectra of the probe wave phase variation on transitions from
the two ground state hyperfine levels (at the same conditions) are shown in
Fig.\ref{expdisp}a. Around $\delta =0$ the phase variations have opposite
slope signs corresponding to opposite signs of the dispersion. The line
shapes are in reasonable agreement with the spectra calculated for
motionless atoms and single closed transitions (Fig.\ref{calcdisp}). This
agreement may seem rather surprising since in the experiment, due to the
velocity distribution, three different atomic transitions, one closed and
two open contribute to the signal in each case. On all open transitions as
well as on the closed $F_g=1\rightarrow F_e=0$ transition EIT occurs and
consequently the dispersion is normal. Only on the $F_g=2\rightarrow F_e=3$
transition EIA takes place and the dispersion is anomalous\cite{Lezama-2}.
However, due to optical pumping the signal is essentially determined by the
closed transitions resulting in the qualitative agreement with the
theoretical prediction. To compare quantitatively the experimental spectra
with theory, velocity distribution, excited state hyperfine splitting and
optical depopulation of open transitions should be taken into account.

The value of $\Delta n=c\delta \Phi /\omega l$ can be obtained taking into
account the cell length ($l=5cm$). The error in the measured absolute value
of the refractive index and the dispersion was around 15\%. However, the
reproducibility of relative measurements was within 2\%.

In the following, we investigate the intensity dependence of the steep
anomalous dispersion. Experimental spectra of the refractive index for
different values of the drive intensity are shown in Fig.\ref{expdisp} b. $%
I_d$ was varied from $0.005\ mW/cm^2$ to $0.05$ $mW/cm^2$ by using filters
while the light diameter in the cell was $1.6$ $cm$. At such low intensities 
$\Delta n$ and $D$ are growing linearly with intensity.{\it \ }In this case,
the refractive index can be characterized by a nonlinear Kerr coefficient $%
n_2$: [$n=n_1+n_2I_d$]. Notice that the value of $n_2$ is a rapidly varying
function of $\delta $. At $\delta =16\ kHz$ we have $n_2\simeq 8\times
10^{-3}\ cm^2/W$. The maximum observed dispersion at low drive intensity was 
$D_{max}\simeq -6\times 10^{-11}Hz^{-1}$.

\begin{figure}[tbp]
\begin{center}
\mbox{\epsfig{file=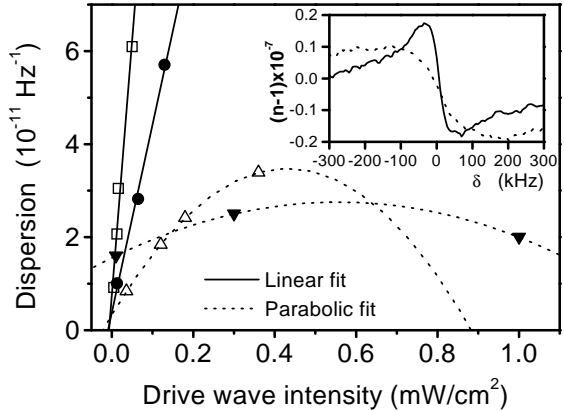,width=3.5in}}
\end{center}
\caption{Measured dispersion for different beam diameters $\varnothing $. Squares: $%
\varnothing =1.6\ cm$, Circles: $\varnothing =1.0\ cm$, Hollow triangles: $%
\varnothing =0.6\ cm$, Solid triangles: $\varnothing =0.4\ cm$. The laser
frequency was locked to the Doppler-free resonance on the $5S_{1/2}$($F_g=2$%
) $\rightarrow $\ $5P_{3/2}$($F_e=3$) transition. Inset: Refractive index as
a function of $\delta $ for $\varnothing =0.3\ cm$ and different
intensities. Solid line:\ $I_d=0.2\ mW/cm^2$, Dotted line:\ $I_d=2\
mW/cm^2$.}
\label{dispdots}
\end{figure}

Based on the previous considerations, the attempt was made to maximize the
dispersion by increasing the drive intensity. Because of laser power
limitation, the increase of $I_d$ was obtained through light cross-section
reduction using a telescope. For three different diameters ($1.6$, $1.0$ and 
$0.6$ $cm$), the drive wave attenuation resulted in dispersion reduction
(Fig.\ref{dispdots}) indicating that we were in the region below the maximum
of the dependence $D(I_d)$. The absolute values of $D$ and $n_2$ were
reduced ($n_2\simeq $ $2\times 10^{-3}\ cm^2/W$ and $n_2\approx 5\times
10^{-4}\ cm^2/W$ for $1.0\ cm$ and $0.6\ cm\ $drive beam diameter,
respectively) in spite of the fact that the intensity was higher. However,
we should notice that the light cross-section reduction results in a
shortening of the interaction time. The observed reduction of the dispersion
for smaller beam diameters indicates that the role of the interaction time
is essential. Only at light wave diameter close to $0.4$ $cm$ we were able
to reach a maximum in the $D(I_d)$ dependence. At relatively high intensity (%
$I_d>1$ $mW/cm^2$) the dispersion decreases with drive intensity and the
refractive index profile becomes significantly different from a dispersion
function (see inset of Fig.\ref{dispdots}). We noticed that within some
range of intensity, $D$ (around $\delta =0$) is almost independent on $I_d$,
while the spectral distance between extrema of $n(\delta )$ is growing with $%
I_d$. This distortion reflects the influence of different time constants.
The precise comparison with theory requires a detailed consideration of the
transient behavior of the coherent medium.

It is possible to characterize a dispersive medium by a light group velocity 
$V_g=c/(n+\nu \times dn/d\nu )$. Steep dispersion ($\mid dn/d\nu \mid \gg
n/\nu $) is intimately associated with large group velocity near the
resonance. The maximum obtained value of steep anomalous dispersion $%
D_{max}\simeq -6\times 10^{-11}Hz^{-1}$ corresponds to a rather slow
negative group velocity $V_g\simeq -c/23000$. It is interesting to mention
that the value of $V_g$ can be infinitely large with opposite sings at
relatively low anomalous dispersion ($dn/d\nu \simeq -$ $n/\nu \simeq
-2.6\times 10^{-15}Hz^{-1}$). According to the linear fits on Fig.\ref
{dispdots} this can be easily obtained with a very weak drive intensity $%
I_d\sim 10^{-5}\ mW/cm^2$. Also in our case a wide range group-velocity
variation can be achieved by switching between normal and anomalous
dispersion (selecting the atomic transition) and/or varying the drive
intensity.

Large dispersion is interesting for different applications. Several ways, in
addition to optimizing $I_d$, can be considered to increase the dispersion
(both normal and anomalous) in degenerate two-level systems. Since $n$ and $%
D $ are proportional to the atomic density while the resonant medium is
thin, it is possible to obtain higher dispersion at higher $N$. In an
optically thick medium, thanks to EIT, further increase of normal dispersion
is possible\cite{Zibrov PRL97}. Another possibility is to use a medium with
very slow ground state relaxation, for instance in a cell with buffer gas 
\cite{50 hz} or a cell with an anti-relaxation coating \cite{Budker-98}.
According to Eq. \ref{refract} one can expect that in this case less
intensity is needed to reach the maximum of the $D(I_d)$ dependence ($\Omega
_{max}^2\sim \gamma \Gamma /2$).

Coherently prepared media with optically controlled dispersion can be of
interest for the design of new devices for pulse delaying/compressing in
communication systems and optical computing. Also, the use of steep
magnetically dependent dispersion for high precision magnetometry was
discussed in detail \cite{Scully}. We should notice, however, that the
requirements on the coherent medium are somehow different for these
applications. For pulse processing elements the bandwidth should be rather
wide to permit operation with short pulses. On another hand higher
sensitivity for low frequency variations of magnetic field require narrow
resonances. Steep dispersion on driven degenerate two-levels systems appear
suitable for the two types of applications.

In conclusion, we have demonstrated for the first time, in agreement with
the theoretical prediction, steep normal and anomalous dispersion in a
driven degenerate two-level atomic system. This result clearly stresses the
importance of degenerate two-level systems for the investigation of quantum
coherence and applications.

This work was supported by CONICYT (Project 92048), CSIC and PEDECIBA
(Uruguayan agencies).

\end{document}